\documentclass[12pt,tightenlines,eqsecnum,floats,floatfix,nofootinbib,amsmath,amssymb,aps,prd,showpacs]{revtex4}

\makeatletter
\let\ftype@table\ftype@figure
\makeatother

\usepackage{color}
\usepackage{graphicx}
\usepackage{tikz}
\usepackage{enumitem}
\usepackage{stmaryrd}

\newcommand{\category}[1]{\bigskip\textit{#1}\enspace}
\newcommand{\pt}[1]{(\textit{#1})}
\newcommand{\be}{\nopagebreak[3]\begin{equation}}
\newcommand{\ee}{\end{equation}}
\newcommand{\bra}[1]{\langle{#1}|}
\newcommand{\ket}[1]{|{#1}\rangle}
\newcommand{\bpl}{\textbf{(}}
\newcommand{\bpr}{\textbf{)}}
\newcommand{\unit}{\,\mathrm}
\newcommand{\E}{\ensuremath{\times10^}}
\renewcommand{\Re}[1]{\operatorname{Re}\,\{{#1}\}}

\newcommand{\dd}{\mathrm{d}}
\renewcommand{\t}{\tilde}
\newcommand{\eps}{\epsilon}

\newcommand{\vx}{{\vec x}}
\newcommand{\vk}{{\vec k}}
\newcommand{\rcr}{\rho_\textrm{max}}
\newcommand{\ton}{t_\textrm{on}}
\newcommand{\tend}{t_\textrm{end}}
\newcommand{\eend}{\eta_\textrm{end}}
\newcommand{\ks}{{k_\star}}
\newcommand{\B}{\textrm{B}}
\newcommand{\BD}{\textrm{BD}}
\newcommand{\I}{\textrm{I}}
\newcommand{\LQC}{\textrm{LQC}}
\newcommand{\Pl}{\textrm{Pl}}
\renewcommand{\today}{0}
\newcommand{\N}{\mathcal{N}}
\newcommand{\Q}{\mathcal{Q}}
\newcommand{\R}{\mathcal{R}}
\newcommand{\T}{\mathcal{T}}
\newcommand{\U}{\mathcal{U}}

\newcommand{\symbolsquare}{\begin{tikzpicture}\fill (0,0) rectangle (0.2,0.2);\end{tikzpicture}}
\newcommand{\symbolcircle}{\begin{tikzpicture}\fill (0,0) circle (0.11);\end{tikzpicture}}
\newcommand{\symboldowntri}{\begin{tikzpicture}\fill (0,0.21) -- (0.24,0.21) -- (0.12,0) -- cycle;\end{tikzpicture}}
\newcommand{\symboldiamond}{\begin{tikzpicture}\fill (0,0.12) -- (0.12,0.24) -- (0.24,0.12) -- (0.12,0) -- cycle;\end{tikzpicture}}
\newcommand{\symbolstar}{\begin{tikzpicture}\fill (0,0.21) -- (0.24,0.21) -- (0.12,0) -- cycle; \fill (0,0.06) -- (0.24,0.06) -- (0.12,0.27) -- cycle;\end{tikzpicture}}

\begin{document}

\title{Detailed analysis of the predictions of loop quantum cosmology for the primordial power spectra}

\author{Ivan Agullo}
\email{agullo@lsu.edu}
\author{Noah A.~Morris}
\email{nmorri7@lsu.edu}
\affiliation{Department of Physics and Astronomy, Louisiana State University, Baton Rouge, LA 70803, USA}

\begin{abstract}

We provide an exhaustive numerical exploration of the predictions of loop quantum cosmology (LQC) with a post-bounce phase of inflation for the primordial power spectrum of scalar and tensor perturbations.
 We extend previous analysis by characterizing the phenomenologically relevant parameter space and by constraining it using observations.   Furthermore, we characterize the shape of LQC-corrections to observable quantities across this parameter space. Our analysis provides a framework to contrast more accurately the theory with forthcoming  polarization data, and it also paves the road for the computation of other observables beyond the power spectra, such as non-Gaussianity.

\end{abstract}


\maketitle

\section{Introduction} \label{s1}

This paper focuses on the quantum gravity extension of the inflationary scenario provided by loop quantum cosmology (LQC) \cite{asrev,agullo-corichi,bojowaldliving,lqcreview}.  We follow the framework introduced by Agullo, Ashtekar and Nelson in a series of papers \cite{aan1,aan2,aan3}. (For other approaches to the early universe in loop quantum cosmology see \cite{calcagni,bojowald&calcagni,barraureview,bbcgk,barrau1,barrau2,barrau3,barrau4,barrau5,wilson-ewing,wilson-ewing2,Cai-Wilson-Ewing,madrid1,madrid2,madrid3,madrid4,reviewmena}.) In short, in this framework quantum gravitational effects dominate the Planck era of the universe, and a quantum bounce appears replacing the classical big bang  singularity. For definiteness, the matter content of the universe is assumed to consist of a massive scalar (``inflaton'') field, although other forms of the potential $V(\phi)$ can be accommodated  without altering the conclusions. Shortly after the bounce, quantum gravity effects gradually lessen and the potential energy $V(\phi)$ begins to prevail. This potential-dominated phase brings the universe, under quite generic circumstances, to a phase of inflation \cite{as,as3}. Therefore, LQC provides an interesting arena to incorporate the highest energy density and curvature stages of the universe into cosmological models, where questions about Planck-scale physics and initial conditions for inflation can be addressed squarely. 

The theory of cosmological perturbations in the Planck era was developed in \cite{aan2}, following previous results in \cite{akl}. It was then applied in \cite{aan3} to show that, generically, the pre-inflationary evolution makes scalar and tensor perturbations reach the onset of inflation in an {\em excited state}, rather than in the Bunch--Davies vacuum often assumed in the cosmology literature. Consequently, the inflationary primordial spectra that source the cosmic microwave background (CMB) anisotropies acquire some extra features with quantum gravitational origin. As described in detail in \cite{aan3} and summarized in section \ref{sec2}, the effects of the pre-inflationary evolution are more important for infrared modes (which correspond to large angles in the CMB), and the size of these effects depends on the parameters of the model: specifically the inflaton mass $m$ and the ``initial'' value at some reference time---for which we choose the bounce time---of the inflaton field, $\phi(t_\textrm{Bounce})\equiv\phi_\B$.

The first phenomenological exploration of the spectrum of scalar and tensor perturbations under this framework appeared in \cite{aan3}. For phenomenological interest, this analysis focused on a small region of the parameter space of the theory, consisting of values of $\phi_\B$ near to its minimum possible value and the inflaton mass $m$ that is commonly used in inflationary cosmology. The main goal of this paper is to extend the study to the \textit{full} parameter space spanning the plane ($\phi_\B,m$).

The goals are multiple: \pt 1 compute the power spectra of scalar and tensor perturbations for the theoretically allowed values of $\phi_\B$ and $m$; \pt 2 identify which region of the ($\phi_\B,m$) plane passes all observational constraints; \pt 3 localize the subspace which, in addition to being observationally allowed, contains significant corrections, originating in the pre-inflationary evolution of LQC, to the standard inflationary picture; and \pt 4 analyze in detail the predicted observational signatures, which will mostly involve tensor modes.

The main challenge of the analysis presented here is computational. Time---and memory---intensive computations using high-performance computing are required to explore the most interesting region this space. But this effort is necessary to understand completely the predictions of the theory across the parameter space. 
 
Our results are in agreement with those obtained previously in \cite{aan3} when we restrict to the range of parameters explored there, but interesting new findings arise in other regions of the parameter space. We summarize here the most important points:
\begin{itemize}
\item Two scales dictate the form of the LQC-corrected power spectrum: First, $k_\LQC$, the momentum scale associated with the spacetime curvature at the bounce, where it attains its maximum value; and second, $k_\I$, the momentum scale associated with the spacetime curvature at the onset of inflation. (It always happens that $k_\LQC \gg k_\I$.)
\item The LQC power spectrum is oscillatory and its average has an amplitude that is amplified with respect to the standard predictions of slow-roll inflation for modes $k_\I \lesssim k \lesssim k_\LQC$, but is in agreement with them otherwise (see Fig.~\ref{full_spectrum}). 
\item The present values of the physical scales $k_\I/a_\today$ and $k_\LQC/a_\today$ ($a_0$ is the scale factor today) depend on the amount of expansion that has occurred since the time of the bounce, which in turn depends on the values of the parameters $\phi_\B$ and $m$. Therefore $\phi_\B$ and $m$ control whether the modes affected by the pre-inflationary dynamics of LQC fall within the window of modes which are observable today. 
\item The region of parameter space which is observationally viable is
    \be \label{region}
    \begin{aligned}
        0.8 \lesssim \phi_\B& \\
        1.1\E{-6}\lesssim m& \lesssim1.5\E{-6} \,.
    \end{aligned}
    \ee
\item The region which, in addition to being observationally allowed, also shows non-negligible LQC modification is an approximately one-dimensional subset of \eqref{region}, which satisfies $\phi_{\B} \approx 1.3\E{-6}/m$ (see Fig.~\ref{map}).
\item LQC corrections to the power spectra tend to make the tensor spectral index $n_t$ more negative and to produce a positive running $\alpha_s$ of the scalar spectral index. 
\item The corrections tend to reduce the the tensor-to-scalar ratio $r$, which serves to alleviate the observational constraints on the $m^2\phi^2$ potential. 
\item The LQC corrections modify the inflationary consistency relation $r/n_t = -8$.
\item The particular choice of initial data for quantum scalar and tensor perturbations has very little impact on the above conclusions, at least for the reasonable choices of initial vacuum state that we have considered in this paper. 
\end{itemize}

In the rest of this paper we provide the details and summarize the calculations leading to these conclusions. We begin in section \ref{sec2} by summarizing  the pre-inflationary physics in LQC, both for the background spacetime and for cosmological perturbations, and by describing the main features of the resulting power spectra for a typical evolution. In section \ref{sec3} we report the results of exploring the predictions of LQC across the parameter space and their relation with observations. Section \ref{sec4} analyzes the sensitivity of the results to the initial quantum state of scalar and tensor perturbations. In section \ref{sec5} we discuss our results and add some final comments. 

Throughout this paper all numerical values are given in Planck units, in which $c=\allowbreak G=\allowbreak \hbar=\allowbreak 1$. Consequently the Planck length, time, and mass all equal unity: $\ell_\Pl=\allowbreak t_\Pl=\allowbreak m_\Pl \,(\equiv\! \sqrt{\hbar c/G})=\allowbreak 1$. However, we will retain $G$ and $\hbar$ explicitly in theoretical expressions to emphasize their physical content.\\
 
\textit{Remark:} In previous analysis \cite{aan1,aan3} WMAP's 7-year observational data was used to compare with observations and to fix some parameters---for instance the inflaton mass. In this paper we use the more recent 2013 \textit{Planck} results \cite{Planck2013CI} ({\em Planck} 2015 results \cite{Planck2015CI} are very similar). Therefore, some care is needed in directly comparing numerical values across the two analyses. 

\section{Summary of the pre-inflationary evolution of loop quantum cosmology} \label{sec2}

The goal of this section is twofold. First, we summarize the physics of the spacetime evolution of the early universe in loop quantum cosmology, as well as the equations governing the evolution of first-order scalar and tensor perturbations thereon. The material has been explained in full detail in the original references \cite{as,as3,aan2,aan3} and summarized in review articles \cite{as,agullo-corichi,ashtekar-barrau}; we therefore provide here only a succinct summary and refer the reader to those references for further subtleties and details. Second, we analyze the main features of the power spectra for a representative concrete choice of initial parameters. This analysis will provide an understanding of the origin and characteristics of the corrections that LQC introduces to the primordial spectra. While some of this material was computed and analyzed in detail in \cite{aan3} as well, we also discuss a number of new features of interest.

\subsection{Dynamics of the FLRW spacetimes in LQC}\label{sec2A}

Since this paper focuses on results directly relevant for observations, we restrict to spatially flat, FLRW spacetimes. The gravitational field is sourced by a single scalar field $\phi$, the inflaton, with an effective potential $V(\phi)$ that we choose to have the simple quadratic form $V(\phi)=\frac12 m^2\phi^2$. As mentioned previously, other choices are certainly possible and, although the concrete numerical values obtained below would change, our findings would remain qualitatively unaltered; the underlying reason for this is that the LQC effects on primordial perturbations originate from quantum-gravitational effects that are largely insensitive to the specific form of $V(\phi)$. 

In LQC the quantum homogeneous and isotropic spacetime is described by a quantum state $\Psi_0(a,\phi)$  which is a complex function of the classical scale factor $a$ (or equivalently its cube, $v=a^3$, which is more commonly used in the LQC literature) and the homogeneous part of the inflaton field  $\phi$. Among the many states  $\Psi_0(a,\phi)$ contained in the LQC Hilbert space \cite{aps1,aps2,aps3,acs}, of particular physical relevance are those which are sharply peaked around a classical trajectory at late times, when the curvature of the universe is well below the Planck scale. Most of these states continue to be sharply peaked during the entire dynamical trajectory \cite{aps2, aps3}, including the Planck era. More importantly, the evolution of the peak of the wavefunction of such states can be accurately described by effective equations that very much resemble the equations of general relativity, apart from some extra terms that account for the quantum gravity corrections [see Eqs.~\eqref{Feq} and \eqref{Req} below]. One can therefore explore the phenomenological consequences of the theory  without the need to solve the complicated discrete equations of the full theory.  For this reason, the effective dynamics have been used in essentially all phenomenological analysis performed so far in the literature. We do the same in this paper.\footnote{One can, however, legitimately argue that other quantum states  $\Psi_0(a,\phi)$ showing large quantum fluctuations in the Planck era may be of physical interest. States $\Psi_0(a,\phi)$ which are not sharply peaked are not accurately described by the effective equations, and to explore their phenomenology one is forced to face the full discrete equations of LQC, which are significantly more complicated. The analysis that we present in this paper has been extended to such more generic family of quantum states, and the results will be presented in a separate publication \cite{aags}.} 
 
The effective equations of LQC were derived in \cite{jw,vt, mb_as_eff1,mb_as_eff2}. The modified Friedmann equations is given by
    \be \label{Feq}
        H^2 = \frac{8\pi G}{3}\rho\bigg(1-\frac{\rho}{\rcr}\bigg) \,;
    \ee
 the modified Raychaudhuri equation is
    \be \label{Req}
        \frac{\ddot a}{a} = -\frac{4\pi G}{3} \rho \bigg(1-4\frac{\rho}{\rcr}\bigg) - 4\pi G\,P  \bigg(1-2\frac{\rho}{\rcr}\bigg) \,;
    \ee
and the equation of motion for the inflaton field, which takes the same form as in the classical theory, is
    \be \label{SFeq}
        \ddot\phi(t) + 3H\,\dot\phi(t) + \frac{\dd V}{\dd\phi}  = 0 \,.
    \ee
In these equations $\rho=\frac12\dot\phi^2+V(\phi)$ and $P=\frac12\dot\phi^2-V(\phi)$ are the energy and pressure density, respectively, of the scalar field; $\rcr=0.41\rho_\Pl$ is the \textit{upper bound} of the energy density in LQC, where $\rho_\Pl$ is the Planck energy density.  Note that $\rcr$ is proportional to $\hbar^{-1}$; thus in the classical limit $\rcr$ diverges and Eqs.~\eqref{Feq} and \eqref{Req} reduce to the classical Friedmann and Raychaudhuri equations.

\category{The relevant parameter space.} The previous equations can be solved numerically by specifying initial data for $a$, $\dot a$, $\phi$ and $\dot\phi$ at a given time, together with a value of the inflaton mass $m$. The bounce is a convenient time to specify initial data because the universality of the solutions there reduces the number of free parameters, as we now explain. First, note that in a flat FLRW spacetime only ratios between values of the scalar factor at different times have objective physical meaning, not the value of $a(t)$ itself at any one time. We are therefore free to rescale $a$ at our convenience. A convenient choice is $a(t_\B)=1$, at the bounce time $t_\B$. Second, note that, because the potential $V(\phi)$ is symmetric, the transformation $\bpl\dot\phi(t_\B), \phi(t_\B)\bpr \longrightarrow \bpl{-\dot\phi(t_\B)},-\phi(t_\B)\mathbf\bpr$ does not alter the physics. We can therefore restrict our solutions to $\dot\phi(t_\B)\geq0$ without loss of generality. Third, at the bounce we always have $\dot a=0$. Fourth, from Eq.~\eqref{Feq} it is easy to see that the value of $\rho$ at the bounce must equal $\rcr$. This implies that $\phi(t_\B)$ and $\dot \phi(t_\B)$ are related by $\rho(t_\B)=\frac12 \dot \phi(t_\B)^2+\frac12 m^2\phi(t_\B)^2=\rcr$. 

In summary, solutions to Eqs.~\eqref{Feq}, \eqref{Req} and \eqref{SFeq} form a two-parameter family  labeled  by couples $\bpl\phi(t_\B), m\bpr$. (From now on we will denote $\phi(t_\B)$ by $\phi_\B$.) Because the energy density is bounded above, the parameters $|\phi_\B|$ and $m$ are as well---they must satisfy $m |\phi_\B| \leq 0.90$. For definiteness, we will assume $\phi_\B>0$ because the sign will not make any qualitative difference in our analysis. We conclude, therefore, that the relevant parameter space for this paper is the set of couples $(\phi_\B,m)$ satisfying 
    \be \label{paramspace}
        0 \leq m\,\phi_\B \leq 0.90 \,.
    \ee

\bigskip

\begin{figure}[tb]
    \includegraphics[width=4.5in]{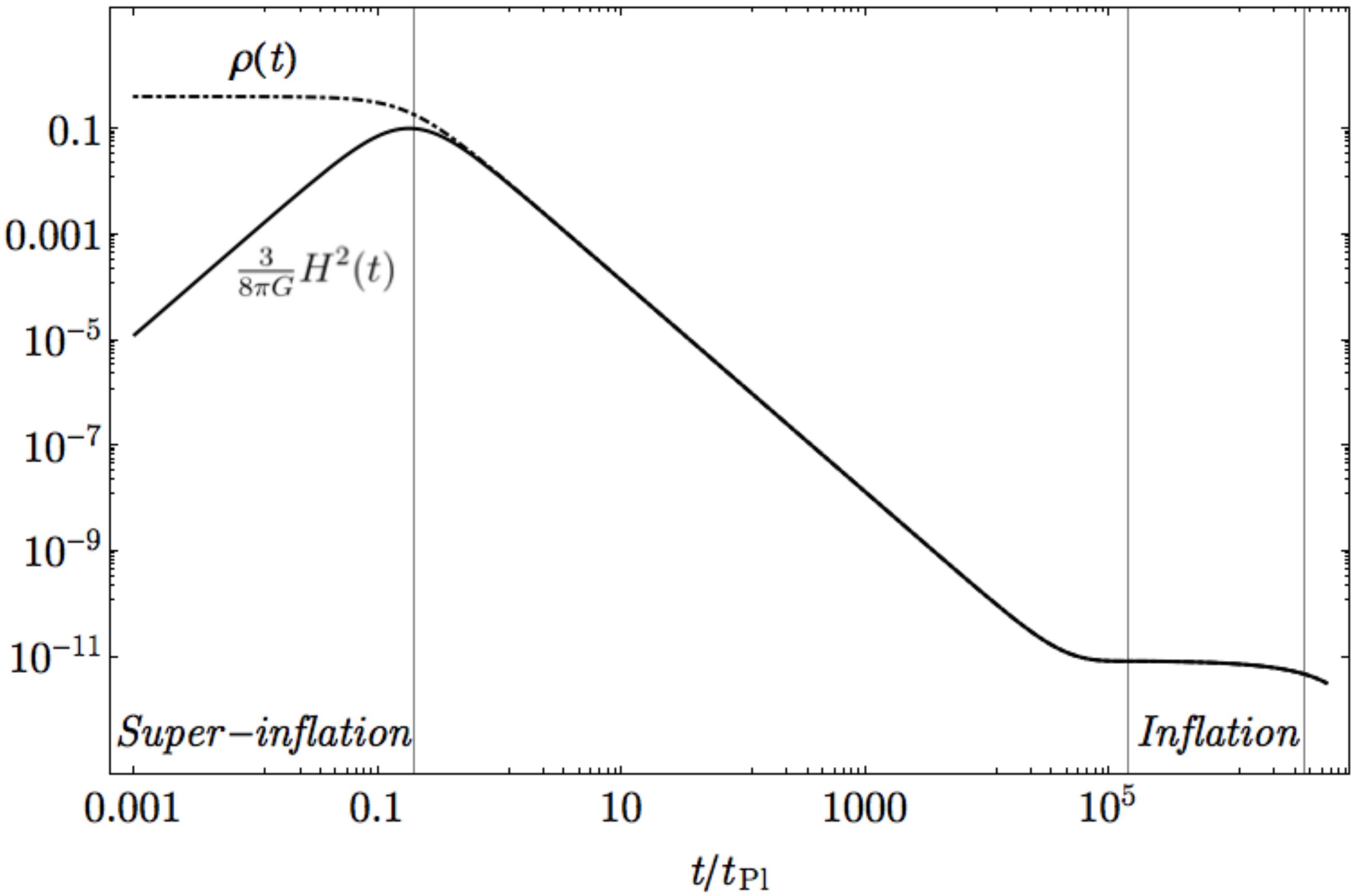}
    \caption{A typical evolution of the energy density and Hubble rate of the background spacetime (in Planck units). Note that both axes are logarithmically scaled. The LQC modification to the classical Friedmann equation is evident near the bounce (since classically $\rho=\frac{3}{8 \pi G}H^2$), but by 2 Planck seconds the behavior coincides with the classical trajectory. \label{background_evolution}}
\end{figure}

We now briefly summarize the main qualitative features of the solutions to Eqs.~\eqref{Feq}, \eqref{Req} and \eqref{SFeq}, which are illustrated in Fig.~\ref{background_evolution}. All solutions experience a bounce at which $H=0$ and $\rho=\rcr$. The bounce has a quantum gravitational origin that makes it independent of the matter content of the theory, and in particular independent of the form of the potential $V(\phi)$. The effective spacetime geometry is symmetric around $t_\B$ for a kinetic-dominated bounce ($\dot \phi(t_\B)^2\gg m^2 \phi_\B^2$), which, as explained in section \ref{sec3}, turns out to be the most interesting regime for possible new predictions. We will therefore focus our discussion on this regime. Immediately after the bounce, the Hubble rate $H$ grows from zero to its maximum value $H_{\max} \approx 0.93$, which is attained around $0.2$ Planck seconds after the bounce. Because $\dot H>0$, this period is commonly known as super-inflation (there is an essentially symmetric period of super-deflation prior to the bounce). After super-inflation, $\dot H$ becomes negative as the inflaton field keeps climbing up the potential at the expense of its kinetic energy, gradually entering a potential-dominated regime. Around $10^5$ Planck seconds after the bounce, the inflaton loses its last remaining kinetic energy, stops moving upward and begins a phase of slow-roll back down the potential; this is the onset of inflation. At this time the energy density has decreased approximately eleven orders of magnitude since the bounce, and consequently the quantum effects of gravity are negligible. The duration of the slow-roll phase depends on how high the inflaton has been able to climb up the potential; this grows monotonically with $\phi_\B$.

In order to define more precisely the times at which the phase of slow-roll inflation begins and ends, we introduce the first-order slow-roll parameters\footnote{The slow-roll parameters defined here, which are calculated in terms of the Hubble parameter, differ from another frequently-encountered set of slow-roll parameters, in terms of the potential $V$: \begin{equation*} \eps_V = \tfrac{1}{2V^2}\big(\tfrac{\partial V}{\partial\phi}\big)^2, \quad \delta_V = \tfrac{1}{V}\tfrac{\partial^2 V}{\partial\phi^2}. \end{equation*} While these two sets of parameters are distinct, they are related if $\dot H$ and $\ddot H$ are sufficiently small by $\eps_V\approx\eps$, $\delta_V\approx\eps-\delta.$}
    \be
        \eps = -\frac{\dot H}{H^2}, \quad \delta = \frac{\ddot H}{2\dot HH}.
    \ee
Slow-roll inflation is said to have begun when both of these parameters are much smaller than unity; for concreteness, we define the onset of slow-roll inflation in this work as the time $\ton$ such that $\eps(\ton) = 0.1$ and $\delta(\ton) = 0.1$, and the end of slow-roll inflation $\tend$ as the first time after $\ton$ when $\eps(\tend) = 1$.

\subsection{Evolution of cosmological perturbations}

In the standard inflationary scenario, cosmic non-uniformities are described by first-order scalar and tensor perturbations. Scalar perturbations can be conveniently described by the Mukhanov--Sasaki variable $\Q(x)$,\footnote{When the matter sector is a single scalar field, this variable relates to the standard comoving curvature perturbation $\R $ by $\R=\frac{H}{\dot\phi}\Q$. As explained in section V of \cite{aan3}, to analyze the pre-inflatioanry universe it is more convenient to use the gauge-invariant variable $\Q$ in place of $\R$, and convert its power spectrum $P_\Q(x)$ to the desired power spectrum $P_\R(x)$ at the end of inflation, than it is to compute the power spectrum of the field $\R(x)$ directly.} and we will collectively denote by $\T(x)$ the two degrees of freedom in tensor perturbations. Because the background energy density is well below the Planck scale during inflation, perturbations are accurately described as quantum fields propagating in the classical FLRW inflationary spacetime. This mathematical framework of quantum field theory in classical but curved spacetimes was developed in the late 1960s and '70s, and since then it has been successfully applied to multiple interesting physical situations. But prior to inflation, and particularly in the vicinity of the bounce, the quantum effects of gravity are no longer negligible, and a description in terms of a classical spacetime metric must be abandoned. One needs to learn how quantum fields propagate in a \textit{quantum} gravitational background.

The quantum theory of cosmological perturbations in the Planck regime was developed in \cite{aan2}, based on previous work by Ashtekar, Kaminski and Lewandowski \cite{akl}, and the result can be summarized as follows. In the regime in which perturbations can be treated as test fields, i.e., their back-reaction on the FLRW spacetime is small, the equations of motion of the operators representing scalar and tensor perturbations are \textit{formally the same} as the equations appearing in classical spacetimes, which in Fourier space are
    \begin{gather} \label{Qeqn}
        \hat\Q_\vk''(\t\eta) + 2 \frac{\t a'}{\t a}(\t\eta)\, \hat\Q_\vk'(\t\eta) + \big(k^2 + \t\U(\t\eta)\big)\hat\Q_\vk(\t\eta) = 0 \,, \\
    \label{Teqn}
        \hat\T_\vk''(\t\eta) + 2
\frac{\t a'}{\t a}(\t\eta)\, \hat\T_\vk'(\t\eta) + k^2\, 
\hat\T_\vk(\t\eta) = 0 \,.
    \end{gather}
A prime indicates a derivative with respect to conformal time $\t\eta$, and $k=|\vk|$. However, the background quantities that appear in these equations, namely $\t a$, $\t\U$ and $\t\eta$, \textit{are different} from their classical counterparts.\footnote{The classical potential that appears in the equation for $\Q_\vk(\eta)$ is $\U(\eta)=a^2 [V(\phi)\,r - 2V_\phi(\phi)\sqrt{r} + V_{\phi\phi}(\phi)]$, where $r=3a^2\phi'^2\,\frac{8\pi G}{\rho}$, $V(\phi)$ is the inflaton potential, and $V_{\phi}(\phi)\equiv \dd V(\phi)/\dd\phi$.} In contrast, $\t a(\eta)$, $\t\U(\eta)$ and $\t\eta$ are now obtained as complicated \textit{expectation values} in the background quantum state $\Psi_0(a,\phi)$, which involve the associated quantum operators $\hat a$ and $\hat\U$ as well as the Hamiltonian background operator (see section II.C of \cite{aan3}). Their explicit form will not be needed for this summary. Remarkably, because of the formal analogy with quantum field theory in classical spacetimes, it is possible to import all the well-developed mathematical machinery from that framework---e.g., Fock quantization, transition amplitudes, renormalization and regularization---to build a well-defined quantum field theory in quantum spacetimes. Therefore, once the quantum backgrounds quantities $\t a$, $\t\U$ have been computed, the evolution of perturbations and computation of observables closely follows the formalism commonly used in semiclassical cosmology. All the difficulties of the quantum spacetime $\Psi_0(a,\phi)$ are encoded in the expectation values $\t a(\t\eta)$ and $\t\U(\t\eta)$. This formalism goes under the name \textit{dressed geometry approach}, because the evolution of perturbations turns out to be mathematically equivalent to the evolution in a curved spacetime whose metric has been \textit{dressed} by quantum gravity effects. The framework has been applied in \cite{aags} to explore the phenomenology of quantum states $\Psi_0(a,\phi)$ which have large fluctuations in the Planck regime and are therefore not semiclassical in any sense.

But a significant simplification to this process appears when $\Psi_0(a,\phi)$ is chosen to be a highly peaked state whose expectation values are well approximated by the solutions of the effective equations \eqref{Feq} and \eqref{Req}. For such highly peaked states, higher moments of $\Psi_0(a,\phi)$ are well approximated by powers of the simplest expectation values. Consequently, explicit computations show that the scale factor $\t a(\t\eta)$ appearing in \eqref{Qeqn} and \eqref{Teqn} reduces to the solution $a(\eta)$ of the effective equations, and the potential $\U(\eta)$ takes the same form as in the classical theory where the time evolution is now dictated by the effective equations of LQC. Therefore, for states $\Psi_0$ that are highly peaked, the computation of power spectra and other physical relevant quantities in LQC follows the same steps as in general relativity, with the only difference being that the evolution of the scale factor $a(\eta)$ and the background inflaton field $\phi(\eta)$ is now replaced by the solution of the effective equations of LQC. 

The quantization of scalar and tensor perturbations now proceeds in the standard way, which we summarize here. For scalar perturbations, one begins by decomposing the field operator $\hat\Q(x)$ as
    \be \label{RepQ}
        \hat\Q(x) = \int\frac{d^3k}{(2\pi)^3}\,\hat\Q_\vk(\eta)\,e^{i\vk\cdot\vx} 
                  = \int\frac{d^3k}{(2\pi)^3}\,\Big[\hat A_\vk\,q_k(\eta) + \hat A^\dagger_{-\vk}\,q^*_k(\eta)\Big]e^{i\vk\cdot\vx} \,,
    \ee
where the functions $q_k(\eta)$, labeled by $k$, form a orthogonal basis of the subspace of ``positive frequency"---more precisely, positive norm---complex solutions to the equation of motion \eqref{Qeqn}. If the basis elements are chosen to satisfy the normalization condition
    \be
        q_k(\eta) q_k'^*(\eta)-q^*_k(\eta) q'_k(\eta) = \frac{i}{a(\eta)^2} \, ,
    \ee
then $\hat A_\vk $ and $\hat A^\dagger_\vk$ satisfy the usual algebra of creation and annihilation operators $[\hat A_\vk, \hat A_{\vk'}] = [\hat A^\dagger_\vk, \hat A^\dagger_{\vk'}] = 0$, $[\hat A_\vk ,\hat A^\dagger_{\vk'}] = \hbar \, (2\pi)^3 \delta(\vk-\vk')$.

The vacuum is the state annihilated by all $\hat A_\vk$, and the symmetric Fock space is the Hilbert space generated by repeatedly operating on this vacuum with creation operators. Eq.~\eqref{RepQ} is then the representation of the operator $\hat\Q$ in this Hilbert space. It is important to emphasize that the definition of vacuum is tailored to the choice of ``positive-frequency'' basis functions $q_k(\eta)$. In maximally symmetric backgrounds such as Minkowski or de Sitter, one can use the spacetime isometries, together with suitable regularity conditions, to single out a preferred basis $q_k(\eta)$ and a preferred vacuum state. But in spacetimes with fewer isometries, e.g., homogeneous and isotropic FLRW backgrounds with arbitrary scale factor $a(\eta)$, there is no canonical vacuum, and consequently the notion of particle is ambiguous. Note that by using a basis of Fourier modes $q_k(\eta)$ that only depend on the length $k$ of the wave vector, rather than its direction, one is already restricting to a family of vacua in which all members are isotropic and homogeneous; this is manifest in the form of the two-point function written below. But there still remains infinite freedom in the choice of vacuum, even within this family.\footnote{The adiabatic condition (see \cite{aan2} for a summary) forces the mode functions $q_k(\eta)$ to approach Minkowski positive-frequency modes at a specific rate when the physical momentum $k/a(t)$ is large compared to the spacetime scalar curvature. But this is an asymptotic condition, and therefore there are infinitely many choices of basis modes $q_k(\eta)$  satisfying it.}

In a free theory, all vacuum correlation functions can be written in terms of the two-point function, which---given a choice of basis $q_k(\eta)$---can be computed as
    \be
        \bra{0} \hat\Q(\eta_1, \vx) \hat\Q(\eta_2, \vx+\Delta\vx) \ket{0} = \hbar \int\frac{d^3k}{(2\pi)^3}\,q_k(\eta_1)q_k^*(\eta_2)\,e^{i\vk\cdot\Delta\vx} \,.
    \ee
The relevant observable in cosmology is the two-point function in momentum space at concurrent times. For a homogeneous and isotropic vacuum, this two-point function is diagonal in the two momenta involved, and a simple computation shows it is given by
    \be
        \bra{0} \hat\Q_\vk(\eta) \hat\Q_{\vk'}(\eta)\ket{0} = (2\pi)^3 \delta(\vk+\vk')\,\frac{2\pi^2}{k^3}P_\Q(k,\eta) \,,
    \ee
where $P_\Q$ is the power spectrum, which is written in terms of the mode functions as $P_\Q(k,\eta) = \hbar \frac{k^3}{2\pi^2}|q_k(\eta)|^2$. Although this power spectrum of the Mukhanov--Sasaki variable is easier to compute, the quantity which is more directly related to observations and thus more interesting in inflationary cosmology is the power spectrum of comoving curvature perturbations $P_\R$. Evaluated at the end of inflation $\eend$, this power spectrum can be obtained from $P_\Q$ as
    \be
        P_\R(k) = \bigg(\frac{H(\eend)}{\dot\phi(\eend)}\bigg)^2 P_\Q(k, \eend) \,.
    \ee

During the inflationary era, sometimes it is useful to write the mode functions $q_k(\eta)$ obtained from the pre-inflationary evolution in terms of the modes $q^\BD_k(\eta)$ that define the Bunch--Davies vacuum during slow-roll inflation
    \be
        q_k(\eta) = \alpha_k\,q_k^\BD(\eta) + \beta_k\,q_k^\BD{}^*(\eta) \,,
    \ee
where $q^\BD_k(\eta) = \sqrt{\eta \pi/4a^2}\, H^{(1)}_\mu(-k\eta)$, with $H^{(1)}_\mu(x)$ a Hankel function of the first kind of order $\mu=3/2+2 \eps+\delta$. Then, $P_\R(k)$ can be written as the power spectrum for the modes $q^{\BD}_k$ times a factor encoding the pre-inflationary evolution which involves the Bogoliubov coefficients $\alpha_k$ and $\beta_k$:
    \be \label{PPBD}
        P_\R(k) = P^{(0)}_\R(k) \, |\alpha_k+\beta_k|^2 \,,
    \ee
where 
    \be \label{PBD}
        P^{(0)}_\R(k) = \bigg(\frac{H(\eend)}{\dot\phi(\eend)} \bigg)^2 \frac{k^3}{2\pi^2}\, \big|q^\BD_k(\eend)\big|^2 
        = \hbar\, \frac{4\pi G}{\eps(\eta_k)} \bigg(\frac{H(\eta_k)}{2\pi}\bigg)^2 \,,
    \ee
with the Hubble exit time for the mode $k$, $\eta_k$, defined by the relation  $k/a(\eta_k)=H(\eta_k)$.

The analysis of tensor perturbations is analogous. The field operator is now expanded in terms of the elements $e_{k}(\eta)$ of a basis of the space of ``positive-frequency'' complex solutions of Eq.~\eqref{Teqn} as
    \be \label{RepT}
        \hat\T(x) = \int \frac{d^3k}{(2\pi)^3}\, \hat\T_\vk(\eta) \, e^{i\vk\cdot\vx}
        = \int \frac{d^3k}{(2\pi)^3}\, \Big[\hat B_\vk \, e_k(\eta) + \hat B^\dagger_{-\vk} \, e^*_k(\eta)\Big]\, e^{i\vk\cdot\vx} \,,
    \ee 
where the basis functions are normalized to 
    \be
        e_k(\eta) e_k'{}^*(\eta)-e^*_k(\eta) e'_k(\eta) = 32\pi G\frac{i}{a(\eta)^2} \, .
    \ee
The power spectrum for each polarization is given by 
    \be \label{PPTBD}
        P_\T(k) = \hbar \frac{k^3}{2\pi^2} |e_k(\eta_{\rm end})|^2
        = P^{(0)}_\T(k)\,  |\alpha^\T_k+\beta^\T_k|^2\, ,
    \ee
where $P^{(0)}_\T(k) = \hbar\, 32\pi G \Big(\frac{H(\eta_k)}{2\pi}\Big)^2$, and $\alpha^{\mathcal{T}}_k$ and $\beta^{\mathcal{T}}_k$ are the Bogoliubov coefficients relating $e_k(\eta)$ and the Bunch-Davies modes $e^{\BD}_k=\sqrt{\eta \pi/4a^2}\, H^{(1)}_\nu(-k\eta)$, with $\nu=3/2+\epsilon$, during slow-roll inflation.

\subsection{The LQC power spectrum}

The scalar and tensor power spectra in LQC were computed and analyzed in detail in \cite{aan3} following the theoretical framework we have just described above (see \cite{calcagni,bojowald&calcagni,barraureview,bbcgk,barrau1,barrau2,barrau3,barrau4,barrau5,wilson-ewing,wilson-ewing2, Cai-Wilson-Ewing, madrid1,madrid2,madrid3,madrid4,reviewmena} for other approaches within LQC). Here, we will conclude this section by summarizing some results found in \cite{aan3} and also presenting some new features.

We have seen that a unique evolution corresponds to each choice of the parameters $(\phi_\B, m)$. For definiteness, we will consider in this section the power spectra generated by choosing $m = 1.3\E{-6}$, which corresponds to the value that is commonly used in standard inflation, and $\phi_\B=1$. (Recall that we use Planck units throughout this paper.) We also must supply concrete initial conditions for the perturbations. As an illustrative example for this section we choose the \textit{preferred instantaneous vacuum} introduced in \cite{ana}; we will impose this vacuum at an initial time $25000$ Planck seconds before the bounce, when all the modes of interest are inside the curvature radius. As previously discussed, and as will be shown explicitly in section \ref{sec4}, other reasonable choices of initial vacua and initial times produce power spectra which are all very similar. 

It is useful to provide a qualitative understanding of the physical evolution of perturbations across the bounce. There are two relevant energy scales in the problem. On the one hand, LQC introduces a {new energy scale} $k_\LQC/a(t_\B) \equiv \sqrt{R_\B/6}\approx 3.21$ that is directly related to the spacetime scalar curvature at the bounce $R_\B=48\pi \rcr\approx 62$. (This is the maximum value that the curvature  attains along its evolution.) A second scale is provided by the value of the scalar curvature at the onset of inflation $k_\I/a(t_\I)\equiv \sqrt{R_\I/6}\approx 10^{-5}$. For kinetic-dominated bounces, the onset of inflation takes place at $t_\I\approx 5\E4$ after the bounce.\footnote{Note that this time is different from the onset of slow-roll, $\ton$, defined previously. Onset of inflation here is defined as the beginning of the phase of accelerated expansion, at which $\eps$ becomes smaller than unity.} To understand the qualitative features of the pre-inflationary evolution of a Fourier mode with co-moving wavenumber $k$, it is convenient to divide the discussion in three different groups:
\begin{itemize}
\item  Fourier modes with co-moving wavenumber $k>k_\LQC$. These modes are ``inside'' the curvature radius at the bounce (i.e., their wavelengths are smaller than the radius of curvature at that time), and will continue to be so until the slow-roll inflationary era. Consequently, spacetime curvature will not affect their evolution until the inflationary era. One expects that these modes will reach the onset of inflation in the Bunch--Davies vacuum, and the final power spectrum will have negligible contributions from the LQC pre-inflationary evolution.
\item Modes with co-moving wavenumber $k_\I<k<k_\LQC$. They are ``outside'' the curvature radius at the bounce. But evolution will bring them inside soon after the bounce, and they will exit again during slow-roll inflation. This process of ``curvature radius-crossing'' enhances the amplitude of the perturbation. In the semi-heuristic language of particle creation, the evolution will create quanta as a consequence of the interaction with spacetime curvature, and the onset of inflation is reached in an excited state. Therefore, we expect the pre-inflationary evolution to affect significantly the power spectra of those perturbations.
\item Modes with co-moving wavenumber $k<k_\I$. They are ``outside'' the curvature radius at the bounce and will continue to be so all the way until the end of inflation. Since these modes do not cross the curvature radius, it is expected that their power spectrum remain small as compared to other modes that do cross it.
\end{itemize}

Therefore, we expect LQC corrections to be relevant for Fourier modes with co-moving wave number $k_\I<k<k_\LQC$. The important question is whether those modes are observable today. To answer this we need to know what physical scales the two scales $k_\LQC$ and $k_\I$ correspond to \textit{at the present time}. The wavenumbers that we can observe in the CMB lie approximately in the interval $(\ks/8.9, 100\ks)$, where $\ks/a_\today=0.002\unit{Mpc^{-1}}$ is the ``pivot'' mode used by the \textit{Planck} satellite team in parameterizing the primordial power spectrum \cite{Planck2013CI}, and corresponds to approximately $\ell=27$ in the angular power spectrum today.\footnote{$\frac{1}{8.9}\ks/a_\today$ equals the Hubble rate today.} Because physical momenta $k/a(t)$ exhibit redshift as the universe expands, the values of the quantities $k_\LQC/a_\today$ and $k_\I/a_\today$ depend on how much expansion has occurred from the bounce and the onset of inflation, respectively, until the present time. The amount of expansion, in turn, is controlled in part by the values chosen for $\phi_\B$ and $m$. Thus the question of whether or not the scale $k_\LQC$ is observable today, and therefore whether the effects of LQC physics may be potentially imprinted in the CMB, will depend on the value of the parameters $\phi_\B$ and $m$. 

\begin{figure}[tb]
    \includegraphics[width=5in]{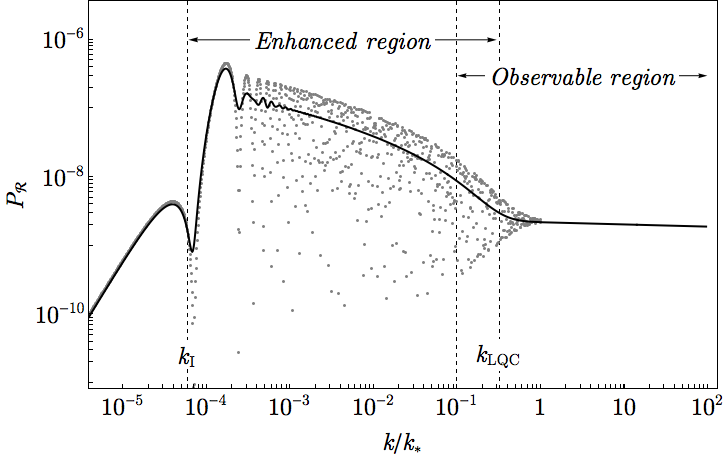}
    \caption{The LQC scalar power spectrum for parameter values $\phi_\B=1$, $m=1.3\E{-6}$, and preferred instantaneous vacuum \cite{ana} initial data for perturbations at initial time $t=-50000$. The numerically evolved spectrum, shown in gray, is rapidly oscillatory; its average, shown in black, has an amplitude  which is amplified with respect to the standard predictions of slow-roll inflation for modes $k_\I \lesssim k \lesssim k_\LQC$ but agrees with them otherwise. \label{full_spectrum}}
\end{figure}

\begin{figure}[tb]
    \includegraphics[width=5in]{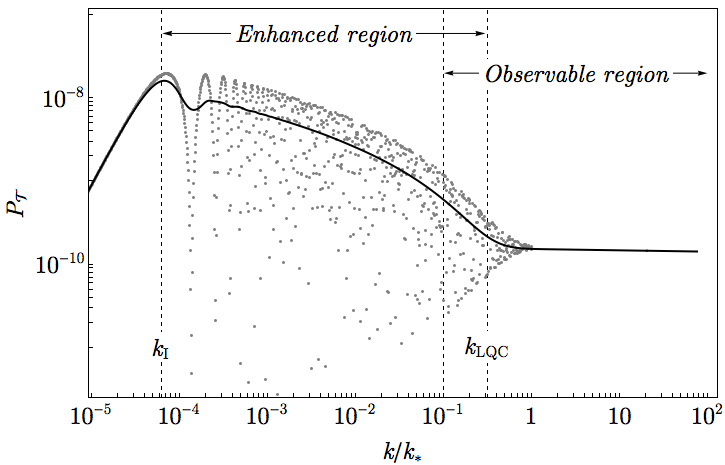}
    \caption{The LQC tensor power spectrum for parameter values $\phi_\B=1$, $m=1.3\E{-6}$, and preferred instantaneous vacuum \cite{ana} initial data for perturbations at initial time $t=-50000$. The numerically evolved spectrum is shown in gray, and  its average in black. The qualitative behaviour is similar to the scalar spectrum shown in Fig.\ \ref{full_spectrum}. \label{full_tensor_spectrum}}
\end{figure}

The above discussion, although very useful to create an intuitive picture, is heuristic. It needs to be supported by explicit computations. Figures \ref{full_spectrum} and \ref{full_tensor_spectrum} show the scalar and tensor power spectrum, respectively, obtained after numerically evolving a large range of co-moving Fourier modes $k$.\footnote{The behavior of the numerically evolved scalar power spectrum is highly oscillatory in $k$; this phenomenon is generically expected for departures from the standard inflationary paradigm (see e.g.\ \cite{brandenberger,easther,danielsson}). All statements we make about the power spectrum should be interpreted as referring to its value after averaging over these high-frequency oscillations, which are so rapid in $k$ as to be unmeasurable by any realistic observation.} As previously mentioned, this figure corresponds to the background values $\phi_\B=1$, $m=1.3\E{-6}$, and  preferred instantaneous vacuum initial data \cite{ana} for perturbations specified in the pre-bounce, contracting phase at $t=-50000$, when all interesting modes are well inside the curvature radius. The figure clearly shows the three regions in $k$-space previously specified. For $k > k_\LQC$ the power spectrum has negligible LQC corrections, and therefore the results agree with the inflationary prediction obtained using  Bunch--Davies vacuum initial data at the onset of inflation. In contrast, the intermediate region $k_\I < k < k_\LQC$ shows a significant enhancement coming from the LQC pre-inflationary evolution. Finally, the power spectrum for the longest wavelengths $k < k_\I$ is largely suppressed. 

For the chosen values $\phi_\B=1$ and $m=1.3\E{-6}$, the LQC scale today $k_\LQC/a_\today$ is around one-third of the pivot scale $\ks/a_\today=0.002\unit{Mpc^{-1}}$. Therefore, the CMB would show some LQC contributions, although only on the largest angular scales $\ell\lesssim 15$. The scale $k_\I/a_\today$ is approximately $10^{-5}$ times $\ks/a_\today$; equivalently, it corresponds to a wavelength $10^4$ times larger than the Hubble radius today. 

To summarize, the LQC corrections to the primordial spectrum of cosmic perturbations are more pronounced for low values of $k$ (i.e., long wavelengths). Depending on the values of the background parameters $\phi_\B$ and $m$, it is possible that LQC corrections begin to appear for the longest-wavelength modes that we can observe, or only for wavelengths that are larger than the Hubble radius today and thus not directly observable. The power spectrum is further amplified for yet larger super-Hubble wavelengths. But remarkably, the power does not grow unboundedly for smaller $k$; on the contrary, it reaches a maximum around $k=k_\I$ and then decreases quite abruptly for lower $k$ values.

The power enhancement of super-Hubble modes in the range between $k_\I$ and $k_\LQC$ is quite interesting and, although counter-intuitive at first, it may lead to additional observable effects. This could be the case if strong \textit{correlations} between observable and super-Hubble modes happen to exist, i.e., if there is considerable non-Gaussianity between these two sets of modes. Under this circumstances, super-Hubble scales can indeed affect the observed power spectrum, introducing a modulation superimposed on the power spectrum we have showed in Fig.~\ref{full_spectrum}. Such modulation could help us to understand the origin of some anomalies discovered by WMAP and confirmed by \textit{Planck} at the largest angular scales of the CMB. The computations of the non-Gaussianity arising as a consequence of the large power spectrum in super-Hubble modes, and its  effects on large angular scales in the CMB, has been analyzed in \cite{a}.

\section{Exploring the parameter space} \label{sec3}

The previous section summarized the theoretical framework needed to propagate cosmic perturbations in the early universe and to compute the power spectra in LQC for a specified choice of initial parameters $\phi_\B$ and $m$. In this section, we consider pairs $(\phi_\B, m)$ throughout the relevant parameter space defined in Eq.~\eqref{paramspace}, and inquire for which such pairs the resulting power spectra are consistent with the strict constraints coming from observations. Then, among this region in the parameter space which is observationally viable, we additionally identify the subset of those points which nonetheless incorporate significant LQC modifications, and we characterize what those modifications are and how they might be distinguished in future observations.

\subsection{Observational constraints} \label{sec3B}

Our ultimate goal is to contrast the result of our computations with observations. To that end, in this section we summarize the constraints stemming from CMB observations that are relevant for our computations. We will  use \textit{Planck} 2013 results \cite{Planck2013CI} for most of our analysis. Observational constraints come entirely from scalar perturbations; for tensor perturbations we have only an upper bound on their amplitude, coming from a joint \textit{Planck}--BICEP2/Keck Array analysis \cite{bkp}: $r(\ks) < 0.12$ ($95\%$ CL). 

\category{1. Amplitude and spectral index of the scalar power spectrum.} By using a phenomenological parametrization of the primordial power spectrum  given by $P_\R(k) = A_s\,(k/\ks)^{n_s-1}$, the \textit{Planck} data, in combination with WMAP and BAO, provide the following values for the best fit of the scalar amplitude $A_s$ and spectral index $n_s$ \cite{Planck2013CI}:
    \be \label{Asns}
        A_s = (2.196^{+0.053}_{-0.058}) \E{-9} \,, 
        \qquad
        n_s = 0.9643 \pm 0.0059 \,.
    \ee
where $1\sigma$ uncertainty ranges are indicated.

\category{2. Running of the spectral index.} When a running $\alpha_s\equiv\dd n_s / \dd\ln k$ is included in the parametrization, $P_\R(k)=A_s\, (k/\ks)^{n_s-1+\frac12 \alpha_s \ln\frac{k}{\ks}}$, the \textit{Planck} 2013 data produce
    \be
        \alpha_s = -0.0013 \pm 0.009,
    \ee
again displaying a $1\sigma$ uncertainty range. This running is compatible with zero and even a positive value at a $1.5\sigma$ level. The inclusion of running does not improve significantly the maximum likelihood of the parametrization.

The recently released new 2015 \textit{Planck} data \cite{Planck2015CI} provide slightly different values for $A_s$, $n_s$ and $\alpha_s$. The impact this change produces on our conclusions is  negligible. 

\category{3. Number of $e$-folds of inflation.} The number of $e$-folds $\N_\star$ between the time $t_\ks$ at which the pivot scale $\ks$ left the Hubble radius during inflation and the end of the inflationary era $\tend$ is constrained \cite{ll2003}. (Recall that $\tend$ is defined here as the time when the slow-roll parameter $\eps=-\dot H/H^2$ reaches the value $\eps=1$ for the first time after slow-roll.) The origin of this constraint is as follows: The pivot scale $\ks$ is, at the present time, $8.9$ times smaller than the Hubble scale, i.e., $H_\today=8.9\, \ks/a_\today$. On the other hand, the Hubble exit time of the mode $\ks$ during inflation, $t_\ks$, is defined by the relation $H(t_\ks)=\ks/a(t_\ks)$. Therefore, by eliminating $\ks$ from these two equations we have a relation between quantities at $t_\ks$ and at the present time: $H_\today \, a_\today/8.9=H(t_\ks)\,  a(t_\ks)$. From  this expression it is straightforward to obtain
    \be
        \N_\star \equiv \ln \frac{a(\tend)}{a(t_\ks)} 
        = \ln 8.9 + \ln\bigg(\frac{a(\tend)}{a_\today}\frac{H(t_\ks)}{H_\today}\bigg) \,.
    \ee
If we make the extreme assumption that the process of reheating is instantaneous and non-dissipative (meaning all energy in the inflaton potential is converted into radiation), then, for the quadratic potential that we use in this paper, $\N_\star$ can be estimated as approximately $61$. A more realistic reheating process could both increase and decrease this quantity; taking into account this uncertainty, we allow the conservative range 
    \be
        50<\N_\star<70 \,.
    \ee

\subsection{Constraining the parameter space}

In this section we report the main computational results of this paper. We have written a numerical code which systematically computes the scalar and tensor power spectra for  values of $\phi_\B$ and $m$ in the relevant parameter space discussed in section \ref{sec2}---pairs $(\phi_\B,m)$ with $0 \leq m\,\phi_\B \leq 0.90$---and contrasts the result against observations. The code is written in \textit{Mathematica}; it evolves the mode functions using \textit{Mathematica}'s native numerical differential-equation solver, which uses an LSODA approach, switching between a non-stiff Adams method and a stiff Gear backward differentiation formula method.  The code was run on the ``Philip'' high-performance computer cluster at Louisiana State University, where it requires approximately 10 processor-hours per point in the parameter space.

Using the observational constraints spelled out in the previous subsection, the code classifies pairs $(\phi_\B,m)$ into the following three (overlapping) categories: \pt 1 Pairs for which the scalar power spectrum is compatible with \textit{Planck} observations for $A_s$ and $n_s$. \pt 2 The subset of the those points  for which the scalar or tensor power spectra contain significant LQC contributions. \pt 3 Points for which the number of $e$-folds $\N_\star$ satisfies the constraint spelled out at the end of previous subsection: $50<\N_\star<70$. In the following we provide further explanation of each category and show the results. 

For definiteness, the computations in this section are done using the preferred instantaneous vacuum initial condition for perturbations imposed at $t=-0.2$.\footnote{We use $t=-0.2$ rather than the bounce $t=0$ because the preferred instantaneous vacuum initial data are not well defined at the bounce time.} Section \ref{sec4} discusses the use of other initial conditions and shows that the results obtained here are unchanged for different reasonable choices of initial vacuum or initial time.

\category{\pt 1 Points $(\phi_\B,m)$ compatible with \textit{Planck} observations for $A_s$ and $n_s$.} Pairs $(\phi_\B,m)$ belong to this category if the scalar power spectrum contains at least one value of $k$ for which the amplitude and tilt agree with Eq.~\eqref{Asns} inside their {\em joint} $1\sigma$ uncertainty region.\footnote{Note that the $1\sigma$ region in the $(A_s,n_s)$ plane is not simply the Cartesian product of the individual $1\sigma$ regions given in Eq.~\eqref{Asns} for $A_s$ and $n_s$. Rather, we model \textit{Planck}'s measured values for the magnitudes and uncertainties of $A_s$ and $n_s$ as following skew-normal and normal, respectively, distributions, and find the (roughly elliptical) contour in the $(A_s,n_s)$ plane for which the \textit{joint} probability is no greater than $1\sigma$.} (The uncertainty ranges make it possible for more than one $k$ to satisfy this condition.) These $k$ values are \textit{candidates} for the pivot scale $\ks$.

\begin{figure}[p!]
    \includegraphics[width=6.3in]{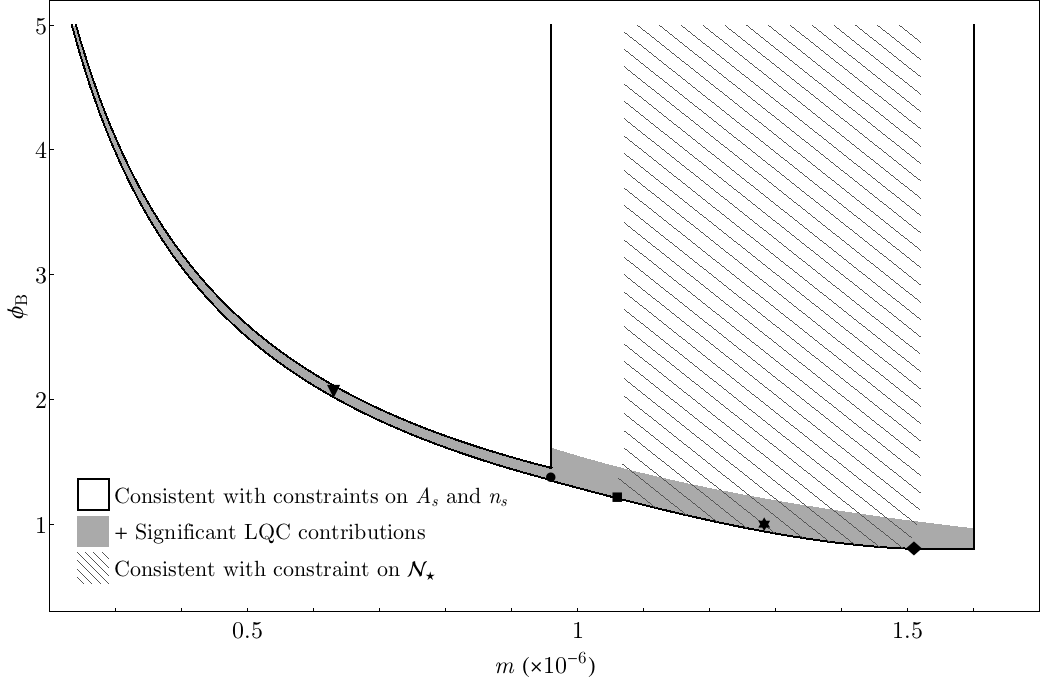}
\caption{Regions of the $(\phi_\B,m)$ parameter space that yield LQC power spectra meeting various criteria. The thick black outline demarcates the points \pt 1 consistent with the \textit{Planck} 2013 observational constraints on the amplitude and tilt of the scalar power spectrum. The gray region \pt 2 indicates the subset of such points for which the spectra contain potentially observable LQC contributions. Finally, the striped region \pt 3 indicates the points consistent with the constraint on the number of $e$-folds $\N_\star$. Several values at selected points from this figure (indicated by the black shapes) are given in Table~\ref{valuetable}.
Due to prohibitive constraints on computational resources, the shapes of the regions \pt 1 and \pt 3 have been extrapolated for the largest shown values of $\phi_\B$. 
\label{map}}
\end{figure}

Points in the plane ($\phi_\B$, $m$) satisfying this condition are in the region of Fig.~\ref{map} outlined by the thick black line, which corresponds to $\phi_\B \gtrsim 0.8$ and $0.9\E{-6}\lesssim m\lesssim 1.6\E{-6}$. Note that these points cover a small portion of the theoretically allowed parameter space. However, that is already the case in the standard inflationary paradigm without LQC, or more generaly in any physical model after contrasting with observational data.

It may seem surprising at first that it suffices to check the values of $A_s$ and $n_s$ at the single mode $\ks$ to ensure compatibility with observations. The reason is that, because the LQC corrections are more important for low $k$'s,  it is guaranteed that pairs $(\phi_\B,m)$ with appropriate amplitude and tilt at $\ks$ also show negligible deviation from a power law for all $k>\ks$. Therefore they are compatible with \textit{Planck} observations.

\begin{table}[p!]
\centering
    \begin{tabular*}{16cm}{@{\extracolsep{\fill}}cccccccccc}
    \hline
    \hline
     & $m$ & $\phi_\B$ & $\ks$ & $r(\ks)$ & $n_t(\ks)$ & $r/n_t$ & $\alpha_s(\ks)$ & $H\bpl\eta_{\ks} \bpr$ & $\eps\bpl\eta_{\ks}\bpr$\\
    \hline
    \symboldowntri & $0.63\E{-6}$ & $2.12$ & $8.3$ & $0.07$ & $-0.035$ & $-2.0$ & $2.5\E{-3}$ & $5.49\E{-6}$ & $0.0044$ \\
    \symbolcircle & $0.96\E{-6}$ & $1.37$ & $10.7$ & $0.11$ & $-0.018$ & $-6.0$ & $3.3\E{-4}$ & $6.70\E{-6}$ & $0.0068$ \\
    \symbolsquare & $1.06\E{-6}$ & $1.22$ & $10.4$ & $0.12$ & $-0.021$ & $-5.8$ & $4.3\E{-4}$ & $7.03\E{-6}$ & $0.0075$ \\
    \symbolstar & $1.28\E{-6}$ & $1.00$ & $16.0$ & $0.14$ & $-0.018$ & $-7.9$ & $2.9\E{-6}$ & $7.81\E{-6}$ & $0.0089$ \\
    \symboldiamond & $1.51\E{-6}$ & $0.79$ & $15.0$ & $0.17$ & $-0.021$ & $-7.9$ & $8.2\E{-6}$ & $8.46\E{-6}$ & $0.0106$ \\
    \hline
    \hline
    \end{tabular*}
    \caption{Values of the tensor-to-scalar ratio $r$, the tensor spectral index $n_t$, the quotient $r/n_t$ appearing in the consistency relation, the running of the scalar spectral index $\alpha_s$, and the value of the Hubble rate $H$ and the slow-roll parameter $\epsilon$ at $\eta_{k^\star}$ for various points $(m,\phi_\B)$ selected from the parameter space. The symbols appearing at the left of the table correspond to the symbols in Fig.~\ref{map} above.
    \label{valuetable}}
\end{table}

\category{\pt 2 Points $(\phi_\B,m)$ compatible with \textit{Planck} observations for $A_s$ and $n_s$ that incorporate significant corrections from LQC.} This category is made of the subset of points from category \pt 1 which show at least a $10\%$ contribution from LQC-physics. To more precisely define  what ``contribution from LQC-physics'' means, it is convenient to write the scalar power spectrum in terms of the auxiliary power spectrum defined in Eq.~\eqref{PBD}; then, the amplitude and tilt can be written as
    \begin{gather}
        \label{NNBD}
        P_\R(k) = P^{(0)}_\R(k) \, |\alpha_k+\beta_k|^2 \,, \\ \nonumber
        n_s(k)-1 = n^{(0)}_s(k)-1 + \frac{\dd\ln|\alpha_k+\beta_k|^2}{\dd\ln k}\, ,
    \end{gather}
where $P^{(0)}_\R(k)$ was given in Eq.~\eqref{PBD} and $n^{(0)}_s(k)-1 = -4\eps(\eta_k) - 2\delta(\eta_k)$. These expressions are very useful because they neatly codify the contribution of pre-inflationary physics into the Bogoliubov coefficients $\alpha_k$ and $\beta_k$, while the standard inflationary contributions are included in $P^{(0)}_\R(k)$ and $n^{(0)}_s(k)$. Therefore, we will say that a given couple  $(\phi_\B,m)$ contains significant contributions from LQC physics when the $\alpha_k$ and $\beta_k$ contributions modify somewhere (in $k$-space) the inflationary results by at least $10\%$. The factor $|\alpha_k+\beta_k|^2 $ can be rewritten, by taking into account the normalization condition $|\alpha_k|^2 - |\beta_k|^2 = 1$, as $1 + 2 |\beta_k|^2 + 2\Re{\alpha_k\beta_k^*}$. We observe in our computations that the interference term $\Re{\alpha_k\beta_k^*}$ is, for the range of $k$'s relevant for this section, highly oscillatory with zero average. Therefore, we average it out and will say that a given couple  $(\phi_\B,m)$ contains significant contributions from LQC physics when 
    \be
        2|\beta_k|^2>0.1
        \qquad \textrm{or} \qquad
        \frac{1}{n^{(0)}_s(k)-1}\frac{\dd\ln|1+2\beta_k|^2}{\dd\ln k} > 0.1
    \ee
for \textit{some} value of $k$ in the observable range.

The pairs ($\phi_\B$, $m$) satisfying this condition appear in gray in Fig.~\ref{map}. They are distributed narrowly around a curve given approximately by $\phi_\B=1.3\E{-6}/m$. For all these points, the LQC corrections appear for the lowest values of $k$. 

Table~\ref{valuetable} indicates the values of the tensor-to-scalar ratio $r$, tensor spectral index $n_t$,  scalar running $\alpha_s$, and other quantities of interest for some points in this category. 
We observe that:
\begin{itemize}
\item The LQC corrections \textit{increase} for lower values of $m$ along the curve $\phi_\B=1.3\E{-6}/m$. In previous analysis \cite{aan3} it was pointed out that the LQC corrections decrease for larger values of $\phi_\B$ when $m$ is held fixed. This is manifest in Fig.~\ref{map} where, for any given value of $m$ in the range $(0.96, 1.6)\E{-6}$, LQC corrections weaken when we move vertically upward in the figure (becoming negligible once out of the gray region). When the freedom in $m$ is included, our results indicate that LQC corrections increase rapidly for lower values of $m$, more than compensating for the effect of increasing $\phi_\B$, when we approximately follow the curve $\phi_\B=1.3\E{-6}/m$ toward smaller $m$.
\item The trend of the LQC-corrections is to make $n_t$ \textit{more negative}, and to \textit{increase} $\alpha_s$. This can be understood by simple inspection of Figs.~\ref{full_spectrum} and \ref{full_tensor_spectrum}, where the corrections increase for lower $k$, therefore making the spectral index more negative and increasing its running. 
\item LQC corrections \textit{decrease} both the value of the slow-roll parameter and the Hubble rate at the time the reference mode $\ks$ exited the Hubble radius during inflation. The reason these values are modified is as follows. The scalar amplitude $A_s$ and spectral index $n_s$---whose values are fixed, up to error bars, by the observational constraints of Eq.~\eqref{Asns}---are given in LQC by\footnote{See Eqs.~\eqref{NNBD}. The latter of these equations has been particularized here for the quadratic potential discussed in this paper; however, the same argument holds for other forms of $V(\phi)$.}
    \be
        \label{Askstar}
        A_s \equiv P_\R(\ks) = \hbar \frac{4\pi G}{\eps(\eta_\ks)} \bigg(\frac{H(\eta_\ks)}{2\pi}\bigg)^2 |\alpha_\ks+\beta_\ks|^2
    \ee
and
    \be
        \label{nskstar}
        n_s(\ks) = 1 - 4\eps(\eta_\ks) + \frac{\dd\ln|\alpha_k+\beta_k|^2}{\dd\ln k}\bigg|_{k=\ks} \,.
    \ee
In standard inflation with vacuum initial conditions, the last term in Eq.~\eqref{nskstar} vanishes, while it is negative in LQC; to maintain the observationally mandated value of $n_s$, the value of $\eps(\eta_\ks)$ must be smaller than in standard inflation. Then, since $|\alpha_k+\beta_k|^2\geq 1$, Eq.~\eqref{Askstar} implies that LQC corrections make $H(\eta_\ks)$ smaller as well.

As a consequence of the decreased value of $H(\eta_\ks)$ in LQC, the energy scale of inflation at the time observable perturbations were generated is reduced.
\item LQC corrections tend to \textit{decrease} $r$. As pointed out in \cite{aan3}, the tensor-to-scalar ratio in LQC is
    \be \label{r}
        r(k) = \frac{2 P_{\T}}{P_{\R}}
        = 16 \, \eps(\eta_k)  \frac{|\alpha^{\T}_k+\beta^{\T}_k|^2}{|\alpha_k+\beta_k|^2}
        \approx 16 \, \eps(\eta_k) \, ,
    \ee
where we have made use of Eqs.~\eqref{PPBD} and \eqref{PPTBD}, and in the last equality we have used the fact that the Bogoliubov coefficients for scalar and tensor perturbations are very similar. This expression looks exactly the same as the result that one would obtain in standard inflation with Bunch--Davies vacuum initial conditions (i.e., without LQC corrections). But we noted above that the corrections decrease the value of $\eps(\eta_\ks)$; thus they in turn decrease the predicted value of $r(\ks)$.

Therefore, LQC helps to alleviate the observational constraint on the $m^2\phi^2$ potential. However, when all observational constraints are  imposed, the corrections on $r$ are small, and this potential still remains close to the border of the $95\%$ CL region. We will have to wait to polarization to know if the $m^2\phi^2$ potential is definitively ruled out. But we emphasize that even if that turns out to be the case, the results of this paper will remain valid for other potentials (e.g., see \cite{gupt-bonga} for the detailed analysis of the Starobinsky potential in LQC).
\item The ratio $r/n_t$ can depart from the standard consistency relation of inflation $r/n_t = -8$; the tendency is for this ratio to become \textit{less negative} with decreasing $m$ (see Table \ref{valuetable}). More precisely, in LQC the consistency relation becomes \cite{aan3}
    \be
        r=-8n_t+\frac{\dd\ln (1+2|\beta_k^\T|^2)}{\dd\ln k} \, .
    \ee
The extra term $\frac{\dd\ln (1+2|\beta_k^\T|^2)}{\dd\ln k}$ is negative in LQC, therefore adding a positive term to the ratio $r/n_t$ (recall $n_t$ is negative).
\end{itemize}

\category{\pt 3 Points $(\phi_\B,m)$ compatible with constraint on $\N_\star$.} In this category we add the conditions on the number of $e$-folds to the points in category 1. As described previously in subsection~\ref{sec3B}, consistency with the present size of $\ks/a_\today$ requires $\ks$ to have exited the Hubble radius between $50$ and $70$ $e$-folds before the end of inflation, $50 < \N_\star < 70$. As explained above, imposing that the amplitude and tilt fall within the joint $1\sigma$ region observed by \textit{Planck} provides a range of candidate values of $\ks$ at each point in category 1. We have marked with stripes in Fig.~\ref{map} the region of points for which at least one of these candidates for $\ks$ also satisfies the constraint on $\N_\star$.

\section{Initial conditions for perturbations} \label{sec4}

To compute the power spectrum, in addition to specifying the background parameters $\phi_\B$ and $m$, one also needs to provide initial conditions for perturbations. This is done by specifying the quantum state of perturbations at some instant of time, at least for the Fourier modes of observational interest. At the practical level, as explained in section~\ref{sec2}, this is achieved by providing initial data for the modes functions $q_k(\eta)$ and $e_k(\eta)$ of scalar and tensor modes, respectively, for the values of $k$ we are interested in. As also discussed in that section, in a generic FLRW spacetime there is no preferred or canonical choice for these initial data. The first choice one needs to make is \textit{when} to specify initial data. Two natural times to impose ``vacuum'' initial conditions are the bounce and the far past (see section III\,C of \cite{aan3} for physical arguments in favor of imposing vacuum initial conditions at the bounce in LQC). But even after a choice of initial time is made, there is still the ambiguity of which ``vacuum'' we choose. 
In this section, we analyze the sensitivity of the observable predictions described in section~\ref{sec3} to different choices. The conclusion will be that different reasonable choices of adiabatic vacua, imposed at the bounce or at some time prior, all produce very similar results for observable modes. Thus the main conclusions of section~\ref{sec3} are robust.  

\subsection{Sensitivity to the choice of vacuum initial conditions at a given time}

\begin{figure}[p]
    \includegraphics[width=4.5in]{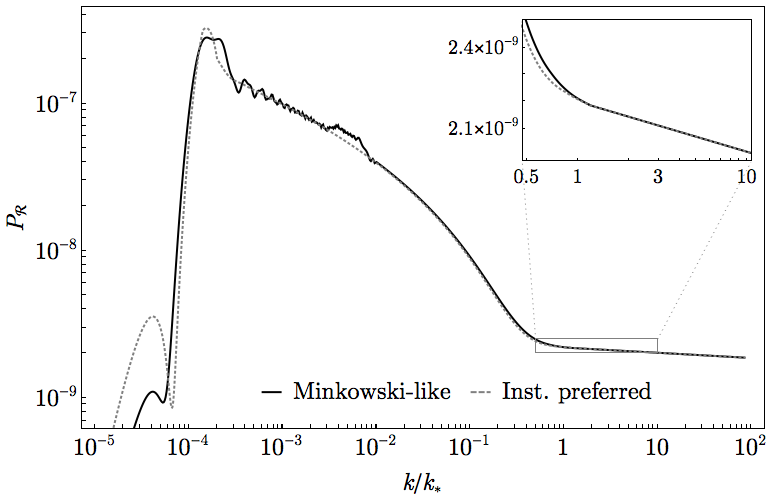}
    \caption{Averaged LQC scalar power spectrum for two different choices of initial data for perturbations specified at $50000$ Planck seconds before the bounce. The two spectra are nearly indistinguishable for observable modes. The background parameters used for both spectra in the figure are $m=1.3\E{-6}$ and $\phi_\B=1$. \label{Power spectra different vacua}}
\end{figure}

\begin{figure}[p]
    \includegraphics[width=5in]{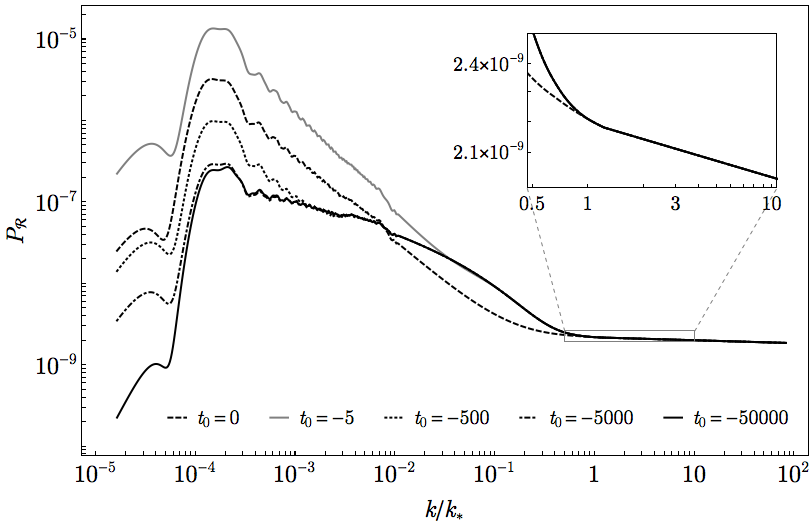}
    \caption{Averaged LQC scalar power spectrum for Minkowski-like initial data specified at different times. $t=0$ corresponds to the bounce, and values of time are given in Planck units. The power spectrum changes significantly for very low values of $k$, indicating that those modes are sensitive to the pre-bounce evolution. On the other hand, all curves are very similar for observable modes, except for the lowest observable values of $k$. The background parameters used for both spectra in the figure are $m=1.3\E{-6}$ and $\phi_\B=1$.\label{spectra different times}}
\end{figure}

We have explored the sensitivity of observable quantities (power spectra and spectral indices) to different choices of vacuum state. At the time of the bounce, our analysis reduces to the one in \cite{aan3}, and our conclusions are the same: observable quantities are quite insensitive to the particular choice of vacuum at the bounce time. We therefore refer the reader to \cite{aan3} for further details. It is important---although perhaps obvious---to emphasize that these conclusions cannot be extrapolated to \textit{any} vacuum state one can write. An arbitrary  Bogoliubov transformation of a given adiabatic vacuum, with appropriate fall-off conditions as $k\to\infty$, is also a legitimate vacuum.  Then, by choosing Bogoliubov coefficients $\beta_k$ appropriately, one can modify the observable quantities arbitrarily much. Our conclusions, as those in \cite{aan3}, are therefore restricted to the small set of states we have explored, which have been selected by demanding extra physical conditions that make those states reasonable candidates for the ground state (see e.g.\ \cite{ana}).

We have extended the analysis in \cite{aan3} to times before the bounce, obtaining similar conclusions. As an example, Fig.~\ref{Power spectra different vacua} shows the scalar power spectrum for two states defined at 50000 Planck seconds before the bounce, namely the preferred instantaneous vacuum introduced in \cite{ana} and the state with Minkowski-vacuum-like\footnote{This state has initial data at $\eta=\eta_0$ given by $q_k(\eta_0)=1/[\,a(\eta_0)\sqrt{2k}\,]$ and  $q'_k(\eta_0)=-ik\, q_k(\eta_0)$ for all $k$. Strictly speaking the resulting state is not adiabatic (it is only zeroth-order adiabatic). However, one can modify these initial data for values of $k$ much larger that the largest $k$ we can observe in order to make the resulting state fourth-order adiabatic; observable quantities are unaffected by this process of ``adiabatization''.} initial data at that time. At times sufficiently before the bounce, all observable modes had physical momentum well above the spacetime curvature scale (i.e., the observable modes were well ``inside'' the curvature radius). In that limit, all vacua of at least second adiabatic order differ only by terms of order $\frac{R(\eta)}{(k/a(\eta))^2}$, where $R(\eta)$ is the Ricci scalar. Therefore, in  that regime different choices of adiabatic vacua all produce very similar result. In the extreme limit $\eta \to -\infty$ there is a preferred notion of ground state: the Minokoski vacuum---or the Bunch--Davies vacuum if a positive cosmological constant is included in the model. 

\subsection{Sensitivity to the choice of time at which vacuum initial conditions are imposed}

We have also explored the extent to which observable quantities are affected by imposing the same notion of vacuum initial data for perturbations at different times.
 Again, we find that predictions 
 are quite insensitive to the specific time chosen  for observable modes. However, we also observe that the power spectrum for modes with smaller wavenumber $k$ (longer wavelength), which today are super-Hubble, are significantly affected. Our result are illustrated in Fig.~\ref{spectra different times}, which  shows the scalar power spectra for $\phi_\B=1$ and $m=1.3\times 10^{-6}$ arising from Minkowski-like initial conditions at different times. The figure shows that the power spectrum for modes $k$ with small ratio $k/k^\star$ changes significantly when initial data is specified at  different times. However, the power spectrum is unchanged for modes with $k/k_\star \gtrsim 1$. The physical reason is that modes with low $k$ exit the curvature radius well before the bounce and consequently their pre-bounce evolution is significantly affected by the spacetime curvature. On the contrary, modes with $k/k_\star \sim 1$ only ``feel'' the spacetime curvature very close to the bounce time, and therefore their power spectrum is insensitive to which time we choose to specify initial data in the contracting phase.

\section{Conclusions} \label{sec5}

In this work we have used observational data to constrain the parameter space of the phenomenological sector of loop quantum cosmology, and we have provided a detailed analysis of the shape of quantum gravitational corrections to observable quantities along this parameter space. We have emphasized that the freedoms in the value of the parameters appearing in the inflaton potential $V(\phi)$ are not fixed by observations alone, and must be included in the parameter space. The computations presented here require the use of high-performance computing. It is our view that this is a necessary task to have full control of the phenomenologically allowed range of parameters and the predictions of the model thereon. We expect our results to be particularly useful for contrasting the theory with the forthcoming data on CMB polarization, as well as to extend the phenomenological explorations in LQC beyond the power spectrum---as for instance in the computations of non-Gaussianity \cite{a}. 

We find particularly remarkable the tight constraints that current observations impose on the parameter space of quantum gravity. When thinking of quantum gravity corrections to observable quantities, one tends to imagine minuscule contributions, hence leaving large freedom for speculation without conflicting with observations. Our analysis shows that this is far from the case. On the contrary,  we have seen that observations strongly constrain the potential effects of quantum gravity. This is more clearly illustrated in Fig.~\ref{map}, where LQC corrections are constrained to a tiny subset of points, distributed in an almost one-dimensional strip, across the theoretically allowed parameter space.

\section*{Acknowledgments}

We thank A. Ashtekar and  B. Gupt for discussions, and J. Tarka for important preliminary work on the project. This work is supported by the NSF Grant No.\ PHY-1403943. Portions of this research were conducted with high performance computational resources provided by Louisiana State University (http://www.hpc.lsu.edu).


\begin{thebibliography}{99}

\frenchspacing 


    
\bibitem{asrev} A. Ashtekar and P. Singh, Loop quantum cosmology: A status report, Class. Quant. Grav. \textbf{28}, 213001 (2011).
    
\bibitem{agullo-corichi} I. Agullo and A. Corichi, ``Loop Quantum Cosmology,'' in \textit{Springer Handbook of Spacetime}, edited by A. Ashtekar and V. Petkov (Springer-Verlag, Berlin, 2014), \texttt{arXiv:1302.3833}.

\bibitem{bojowaldliving} M. Bojowald, Loop quantum cosmology, Living Rev. Relat. \textbf{11}, 4 (2008).    
    
\bibitem{lqcreview} K. Banerjee, G. Calcagni and M. Mart\'in-Benito, Introduction to loop quantum cosmology, SIGMA \textbf{8}, 016 (2012).
    
 \bibitem{aan1} I. Agullo, A. Ashtekar and W. Nelson, A quantum gravity extension of the inflationary scenario, Phys. Rev. Lett. \textbf{109}, 251301 (2012).
  
\bibitem{aan2} I. Agullo, A. Ashtekar and W. Nelson, An extension of the quantum theory of cosmological perturbations to the Planck era, Phys. Rev. D \textbf{87}, 043507 (2013).

\bibitem{aan3} I. Agullo, A. Ashtekar and W. Nelson, The pre-inflationary dynamics of loop quantum cosmology: Confronting quantum gravity with observations, Class. Quant. Grav. \textbf{30}, 085014 (2013).

\bibitem{calcagni} G. Calcagni, Observational effects from quantum cosmology, Annalen Phys. \textbf{525}, no. 5, 323 (2013); \textbf{525}, no. 10--11, A165(E) (2013). 

\bibitem{bojowald&calcagni} M. Bojowald, G. Calcagni and S. Tsujikawa,  Observational test of inflation in loop quantum cosmology, JCAP \textbf{1111}, 046 (2011). 

\bibitem{barraureview} A. Barrau, T. Cailleteau, J. Grain and J. Mielczarek. Observational issues in loop quantum cosmology, Class. Quant. Grav. \textbf{31}, 053001 (2014).

\bibitem{bbcgk} A. Barrau, M. Bojowald, G. Calcagni, J. Grain and M. Kagan, Anomaly-free cosmological perturbations in effective canonical quantum gravity, JCAP \textbf{1505}, 051 (2015).

\bibitem{barrau1} J. Grain and A. Barrau, Cosmological footprints of loop quantum gravity, Phys. Rev. Lett. \textbf{102}, 081301 (2009).
        
\bibitem{barrau2} J. Grain, T. Cailleteau, A. Barrau and A. Gorecki, Fully loop-quantum-cosmology-corrected propagation of gravitational waves during slow-roll inflation, Phys. Rev. D \textbf{81}, 024040 (2010).

\bibitem{barrau3} J. Mielczarek, T. Cailleteau, J. Grain and A. Barrau, Inflation in loop quantum cosmology: Dynamics and spectrum of gravitational waves, Phys. Rev. D \textbf{81}, 104049 (2010).

\bibitem{barrau4} J. Grain, A. Barrau, T. Cailleteau and J. Mielczarek, Observing the big bounce with tensor modes in the cosmic microwave background: Phenomenology and fundamental LQC parameters, Phys. Rev. D \textbf{82}, 123520 (2010).
    
\bibitem{barrau5} T. Cailleteau, J. Mielczarek, A. Barrau and J. Grain, Anomaly-free scalar perturbations with holonomy corrections in loop quantum cosmology, Class. Quant. Grav. \textbf{29}, 095010 (2012).

\bibitem{wilson-ewing} E. Wilson-Ewing, Holonomy corrections in the effective equations for scalar mode perturbations in loop quantum cosmology, Class. Quant. Grav. \textbf{29}, 085005 (2012).

\bibitem{wilson-ewing2} E. Wilson-Ewing, Lattice loop quantum cosmology: scalar perturbations, Class. Quant. Grav. \textbf{29}, 215013 (2012); The matter bounce scenario in loop quantum cosmology, JCAP \textbf{1303}, 026 (2013).    

\bibitem{Cai-Wilson-Ewing} Y. Cai and E. Wilson-Ewing, A $\Lambda$CDM bounce scenario, JCAP \textbf{1503}, 006 (2015).

\bibitem{madrid1} L. Castell\'o Gomar, M. Fern\'andez-M\'endez, G. A. Mena Marug\'an and J. Olmedo, Cosmological perturbations in hybrid loop quantum cosmology: Mukhanov--Sasaki variables, Phys. Rev. D \textbf{90}, 064015 (2014). 
 
\bibitem{madrid2} M. Fern\'andez-M\'endez, G. A. Mena Marug\'an and J. Olmedo, Hybrid quantization of an inflationary universe, Phys. Rev. D \textbf{86}, 024003 (2012).

\bibitem{madrid3} M. Fern\'andez-M\'endez, G. A. Mena Marug\'an and J. Olmedo, Hybrid quantization of an inflationary model: The flat case, Phys. Rev. D \textbf{88}, 044013 (2013).

\bibitem{madrid4} M. Fern\'andez-M\'endez, G. A. Mena Marug\'an and J. Olmedo, Effective dynamics of scalar perturbations in a flat Friedmann--Robertson--Walker spacetime in loop quantum cosmology, Phys. Rev. D \textbf{89}, 044041 (2014).  
 
\bibitem{reviewmena} G. A. Mena Marug\'an, Loop quantum cosmology: A cosmological theory with a view, J. Phys. Conf. Ser. \textbf{314}, 012012 (2011); A brief introduction to loop quantum cosmology, AIP Conf. Proc. \textbf{1130}, 89 (2009).

\bibitem{as} A. Ashtekar and D. Sloan, Loop quantum cosmology and slow roll inflation, Phys. Lett. B \textbf{694}, 108 (2010).

\bibitem{as3} A. Ashtekar and D. Sloan, Probability of inflation in loop quantum cosmology, Gen. Rel. Grav. \textbf{43}, 3619 (2011).

\bibitem{akl} A. Ashtekar, W. Kaminski and J. Lewandowski, Quantum field theory on a cosmological, quantum space-time, Phys. Rev. D \textbf{79}, 064030 (2009).

\bibitem{Planck2013CI} P. A. R. Ade \textit{et al.} (Planck Collaboration), \textit{Planck} 2013 results. XXII. Constraints on inflation, Astron. Astrophys. \textbf{571}, A22 (2014).

\bibitem{Planck2015CI} P. A. R. Ade \textit{et al.}  (Planck Collaboration), \textit{Planck} 2015 results. XX. Constraints on inflation, \texttt{arXiv:1502.02114}.

\bibitem{ashtekar-barrau} A. Ashtekar and A. Barrau, Loop quantum cosmology: From pre-inflationary dynamics to observations, \texttt{arXiv:1504.07559}.  

\bibitem{aps1} A. Ashtekar, T. Pawlowski and P. Singh, Quantum nature of the big bang, Phys. Rev. Lett. \textbf{96}, 141301 (2006).

\bibitem{aps2} A. Ashtekar, T. Pawlowski and P. Singh, Quantum nature of the big bang: An analytical and numerical investigation, Phys. Rev. D \textbf{73}, 124038 (2006).
    	
\bibitem{aps3} A. Ashtekar, T. Pawlowski and P. Singh, Quantum nature of the big bang: Improved dynamics, Phys. Rev. D \textbf{74}, 084003 (2006).	

\bibitem{acs} A. Ashtekar, A. Corichi and P. Singh, Robustness of predictions of loop quantum cosmology, Phys. Rev. D \textbf{77}, 024046 (2008).     
  
\bibitem{aags} I. Agullo, A. Ashtekar and B. Gupt, LQC phenomenology from non-semiclassical quantum states, to appear. 

\bibitem{jw} J. L. Willis, On the low-energy ramifications and a mathematical extension of loop quantum gravity. Ph.D. thesis, The Pennsylvania State University (2004).

\bibitem{vt} V. Taveras, Corrections to the Friedmann equations from LQC for a universe with a free scalar field, Phys. Rev. D \textbf{78}, 064072 (2008).

\bibitem{mb_as_eff1} M. Bojowald and A. Skirzewski, Effective theory for the cosmological generation of structure, Rev. Math. Phys. \textbf{18}, 713 (2006).

\bibitem{mb_as_eff2} M. Bojowald, B. Sandh\"ofer, A. Skirzewski and A. Tsobanjan, Effective constraints for quantum systems, Rev. Math. Phys. \textbf{21}, 111 (2009).

\bibitem{ana} I. Agullo, W. Nelson and A. Ashtekar, Preferred instantaneous vacuum for linear scalar fields in cosmological space-times, Phys. Rev. D \textbf{91}, 064051 (2015).   

\bibitem{brandenberger} R. H. Brandenberger, Introduction to early universe cosmology, in proceedings of ``4th International Conference on Fundamental Interactions'' PoS(ICFI2010)001; R. H. Brandenberger and J. Martin, Trans-Planckian issues for inflationary cosmology, Class. Quant. Grav. \textbf{30}, 113001 (2013).

\bibitem{easther} R. Easther, B. R. Greene, W. H. Kinney and G. Shiu, A generic estimate of trans-Planckian modifications to the primordial power spectrum in inflation, Phys. Rev. D \textbf{66}, 023518 (2002); Imprints of short distance physics on inflationary cosmology, Phys. Rev. D \textbf{67}, 063508 (2003); Inflation as a probe of short distance physics, Phys. Rev. D \textbf{64}, 103502 (2001).

\bibitem{danielsson} U. H. Danielsson, A note on inflation and transplanckian physics, Phys. Rev. D \textbf{66}, 023511 (2002).

\bibitem{a} I. Agullo, Loop quantum cosmology, non-Gaussianity and CMB anomalies, Phys. Rev. D (in press), arXiv:1507.04703. 

\bibitem{bkp} P. A. R. Ade \textit{et al.} (BICEP2 and Planck Collaborations), Joint Analysis of BICEP2/\textit{Keck?Array} and \textit{Planck} Data, Phys. Rev. Lett. \textbf{114}, 101301 (2015).
  
\bibitem{ll2003} A. R. Liddle and S. M. Leach, How long before the end of inflation were observable perturbations produced?, Phys. Rev. D \textbf{68}, 103503 (2003).

\bibitem{gupt-bonga} B. Bonga and B. Gupt, to appear.


\end{thebibliography}
\end{document}